\documentclass[english,aps,preprint,prc]{revtex4}
\usepackage[T1]{fontenc}
\usepackage[latin9]{inputenc}
\usepackage{amsmath}
\usepackage{amssymb}
\usepackage{graphicx}
\def\be{\begin{equation}}
\def\ee{\end{equation}}
\makeatletter

\@ifundefined{textcolor}{}
{%
 \definecolor{BLACK}{gray}{0}
 \definecolor{WHITE}{gray}{1}
 \definecolor{RED}{rgb}{1,0,0}
 \definecolor{GREEN}{rgb}{0,1,0}
 \definecolor{BLUE}{rgb}{0,0,1}
 \definecolor{CYAN}{cmyk}{1,0,0,0}
 \definecolor{MAGENTA}{cmyk}{0,1,0,0}
 \definecolor{YELLOW}{cmyk}{0,0,1,0}
}

\usepackage{babel}

\makeatother

\usepackage{babel}
\begin{document}

\preprint{Draft 2}

\title{Electromagnetic transition strengths in soft deformed nuclei}

\author{L.M. Robledo}

\email{luis.robledo@uam.es}

\homepage{http://gamma.ft.uam.es/robledo}

\selectlanguage{english}%

\affiliation{Departamento de F\'\i sica Te\'orica, M\'odulo 15, Universidad Aut\'onoma de
Madrid, E-28049 Madrid}

\author{G.F. Bertsch}

\email{bertsch@uw.edu}

\selectlanguage{english}%

\affiliation{Institute for Nuclear Theory and Dept. of Physics, Box 351560, University
of Washington, Seattle, Washington 98915, USA}
\begin{abstract}
Spectroscopic observables such as electromagnetic transitions strengths
can be related to the properties of the intrinsic mean-field wave
function when the latter are strongly deformed, but the standard rotational
formulas break down when the deformation decreases.  Nevertheless there 
is a well-defined, non-zero, spherical limit that
can be evaluated in terms of overlaps of mean-field intrinsic deformed
wave functions. We examine the transition between the spherical limit and strongly
deformed one for a range of nuclei comparing the two limiting formulas
with exact projection results.  We find a simple criterion for the validity
of the rotational formula depending on $\langle \Delta \vec{J}^2\rangle$, the
mean square fluctuation in the angular momentum of the intrinsic state.
We also propose an interpolation formula which describes the transition
strengths over the entire range of
deformations,
reducing to the two simple expressions in the appropriate limits.
\end{abstract}
\maketitle

\section{introduction}

In mean-field theories, electromagnetic transition rates are often 
evaluated using the rotational formula\cite{BM} to relate them to the multipole
moments of the mean-field wave functions. 
The formula is justified by factorizing the wave function as 
a product of a wave function for the orientation angles times an intrinsic
wave function and assuming that the matrix elements between intrinsic
states at different orientations vanish.
From a more microscopic
point of view, the formula can be obtained as the strong deformation
limit of the transition probability computed with angular momentum
projected wave functions \cite{man75,RS}. There are several studies
in the literature investigating the validity of the rotational formula
in well deformed nuclei \cite{man75,isl79,rin82}.  However, as far as we
know there has never been a systematic study of the validity and
eventual breakdown of the rotational formula as the wave function approaches
the spherical limit.  A motivation for this study is 
the wide-spread use of this formula even outside of its domain of validity.
For example, the
increasing popularity of the Bohr Hamiltonian \cite{5DBH} as a tool
to handle low energy vibrational and rotational properties in a 
mean-field framework calls for a careful analysis of the limitations
of the rotational formula for $B(E2)$ transition strengths\cite{wil56}. Often 
near-spherical configurations have a non-negligible amplitude in the
wave functions and their contribution to the transition strengths
needs to be handled with care.  The purpose of this paper is to 
establish criteria for the use of rotational formulas, as well as to
find useful approximations simpler than the full angular momentum 
projection to deal with moderate and soft deformations.

This paper is organized as follows.  Sect. II below discusses the
representation of the wave function at small deformations.
Our main result, derived in Sect.III,  is an
an expression for the transition strengths valid for small deformations
Eq. (\ref{eq:BELsph}) below.   This expression gives a 
non-zero value in the limit of vanishing deformation, in contrast 
with the rotational formula, Eq. (\ref{eq:Rot}) below.
In Sect. IV we examine the validity of the formulas by comparing with 
full projections from the intrinsic states, taking
a number of representative examples
including quadrupole and octupole transitions.  
The dividing line separating
the small and large deformation limits is seen to be closely connected
to the the angular momentum content of the intrinsic wave function.  
This gives a simple criterion to identify the
regions of validity of the rotational formula.
We also find that the
$B(E2)$ values can be simply parameterized as a function of the
the quadrupole deformation parameter, Eq. (\ref{eq:BE2ProjAPP}) below.  
Other transition strengths like the $B(E3,3^{-}\rightarrow0^{+})$
will be discussed and we will see that similar considerations apply to 
them as well.

To set the notation, the multipole operators are defined as \cite{Qdef}
\begin{equation}
\hat Q_{\lambda\mu}=\sqrt{\frac{4\pi}{2\lambda+1}}r^{\lambda}Y_{\lambda\mu}
\end{equation}
and the corresponding electric operators  as
\begin{equation}
\hat Q_{\lambda\mu}^e = e\frac{(1-2 \tau_z)}{2}\hat Q_{\lambda\mu}.
\end{equation}
%The $B(EJ; J \rightarrow 0)$ transition strength is then given
%by 
%\begin{equation}
%B(EJ; J \rightarrow 0) = {2 J + 1 \over 4 \pi} |\langle J|\hat Q_{J0}|0\rangle|^2.
%\end{equation}
The rotational formula for an axially symmetric intrinsic state is 
given by 
\begin{equation}
\label{eq:Rot}
B(EJ; J \rightarrow 0)_{\textrm{ROT}} ={1\over 4 \pi} |\langle\phi|\hat
Q_{J0}^e|\phi\rangle|^2.
\end{equation}

\section{Mean field wave functions near sphericity}

The first step is the characterization of the intrinsic wave functions
near sphericity. We will focus on quadrupole deformation because the
generalization to other multipolarities is straightforward. We assume
that the intrinsic wave functions are of the Hartree-Fock-Bogoliubov
(HFB) mean-field type.  The wave function $|\phi(q)\rangle$ is labeled
by the components of the quadrupole moment 
$q_{2\mu}=\langle \phi |\hat Q_{2\mu}| \phi \rangle$ ($\mu=-2,\ldots,2$).  The wave
function can be expressed 
in terms of a suitable spherical reference state $|\phi(0)\rangle$
by means of the generalized Thouless theorem 
\[
|\phi(q)\rangle= \mathcal{N}_{q}\exp(i\hat{Z}(q)|\phi(0)\rangle.
\]
Here $\hat{Z}(q)$ is a sum of 2-quasiparticle creation operators and
$\mathcal{N}_{q}$ is a normalization constant.
Given the Bogoliubov amplitudes $U(q)$, $V(q)$
and $U(0)$, $V(0)$ defining $|\phi(q)\rangle$ and $|\phi(0)\rangle$
(see \cite{RS} for notation) the explicit form of $\hat{Z}(q)$
can be obtained \cite[App. E.3]{RS}. However, we only need to assume for
the formal development below that $\hat Z$ can be expanded as a power
series in $q$,
\[
\hat{Z}(q)=\sum_{\mu}q_{2\mu}(-1)^{\mu}\hat{Z}_{2,-\mu}+\frac{1}{2}\sum_{\mu,\mu'}q_{2\mu}q_{2\mu'}(-1)^{\mu+\mu'}\hat{Z}_{2,-\mu,-\mu'}'+\cdots
\]
 The phases are introduced for consistency with the following properties
of the deformation parameters $q_{2\mu}=
\langle \hat Q_{2\mu}\rangle=
\langle \hat Q_{2\mu}\rangle^*=
(-1)^{\mu}\langle \hat Q_{2-\mu}\rangle=
(-1)^{\mu}q_{2-\mu}$.
It also implies that $\hat{Z}_{2,\mu}^{(1)\,+}=(-1)^{\mu}\hat{Z}_{2,-\mu}$
and $\hat{Z}_{2,\mu,\mu'}^{(2)\,+}=(-1)^{\mu+\mu'}\hat{Z}_{2,-\mu,-\mu'}'$.
%The operators $r^{2}Y_{2\mu}$ constitute the five components of an
%spherical tensor of rank 2. 
The tensor character of the multipole
operators implies that the deformation parameters of the rotated wave
function $|\phi(q'_{2\mu})\rangle=\hat{R}(\Omega)|\phi(q{}_{2\mu})\rangle$
also behave as the components of a spherical tensor $q'_{2\mu}=
\sum_{\mu'}\mathcal{D}_{\mu\mu'}^{2\,*}(\Omega)q_{2\mu'}.$
To be consistent with this property, the operator $\hat{Z}_{2,\mu}$
must transform under rotations as
$$
\hat{R}\hat{Z}_{2\mu}\hat{R}^{+}=\sum_{\mu'}\mathcal{D}_{\mu'\mu}^{2}(\Omega)\hat{Z}_{2\mu'}.
$$
 The corresponding transformation properties of the operators $\hat{Z}_{2,-\mu,-\mu'}'$
are given by 
$$
\hat{R}\hat{Z}_{2,\mu,\mu'}'\hat{R}^{+}=\sum_{\nu\nu'}\mathcal{D}_{\nu\mu}^{2}(\Omega)\mathcal{D}_{\nu'\mu'}^{2}(\Omega)\hat{Z}_{2,\nu,\nu'}'
$$
 This property makes it possible to decompose the operator as the
direct sum of spherical tensors 
$$
\hat{Z}_{2,\mu,\mu'}'=\sum_{JM}\langle2\mu2\mu'|JM\rangle\hat{Z}_{JM}'
$$
In the present example the range of the spherical tensors $\hat{Z}_{JM}'$
is $J=0,\ldots,4$. Using the same kind of arguments it is easy to
show that the $\hat{Z}$and $\hat{Z}'$ operators must
be even under parity. The generalization to an arbitrary multipolarity
$\lambda$ is straightforward; we consider the case $\lambda=3$ in 
more detail below.

\section{Transition strengths in the spherical limit}

Close to the spherical limit, the deformation parameters of the intrinsic
wave function are small and we can expand
$|\phi(q)\rangle$ to second order in $q_{2\mu}$. The wave function
is then projected on good angular momentum using the projection operator
\begin{equation}
\hat P^J_{MK} = { 2J+1\over 8\pi^2}\int d \Omega {\cal D}^J_{MK}(\Omega)
\hat R_\Omega
\end{equation}
and the transformation properties of the $Z$ operators. The ground state
$|0^{+}\text{\ensuremath{\rangle}}$ is obtained by projecting with
$\hat P^0_{00}$.  It is given up to second order in $q_{2\mu}$ by
\begin{equation}
|0^{+}\rangle=\mathcal{N}_{0}\left\{ |\phi(0)\text{\ensuremath{\rangle}}+
q_2^{2}\left(
[\hat{Z}\otimes\hat{Z}]_{0}^{0}
+\frac{1}{2}\hat{Z}_{00}'\right)|\phi(0)\text{\ensuremath{\rangle}}+\ldots\right\} \label{eq:GS0}
\end{equation}
Here we have introduced the notation 
$q_{2}^{2}=\frac{1}{\sqrt{5}}\sum_{\mu}q_{2\mu}q_{2-\mu}(-1)^{2-\mu}$ and 
\begin{equation}
[\hat{Z}\otimes\hat{Z}]_{M}^{J} =\sum_{\mu,\mu'}
\langle 2\mu 2\mu'|JM \rangle \hat{Z}_{2,\mu}\hat{Z}_{2,\mu'}
\end{equation}
Only the first term in Eq. (1), zeroth order in $q_{2,\mu}$, will be required in
the derivations below.
The projection on $J=2$ with the operator $\hat P^2_{MM}$ gives
the excited  $|2^{+}M\rangle$ 
state as
\begin{equation}
|2^{+}M\rangle=\mathcal{N}_{2M}\left\{ (-1)^{M}q_{2-M}\hat{Z}_{2M}|\phi(0)\rangle+O(q_{2M}^{2})\right\} \label{eq:EX2}
\end{equation}
with a normalization factor $\mathcal{N}_{2M}$
given by 
$$
1=|\mathcal{N}_{2M}|^{2}\left(q_{2-M}^{2}\langle\phi(0)|\hat{Z}_{2M}^{+}\hat{Z}_{2M}|\phi(0)\rangle+O(q_{2M}^{3})\right).
$$
Since$|\phi(0)\rangle$ is a spherical wave function, the state $\hat{Z}_{2M}|\phi(0)\rangle$
has angular momentum 2 and the mean value on the right hand
side of the above equation is independent of $M$. It will be written
as $\langle||\hat{Z}_{2}^{+}\hat{Z}_{2}||\rangle$ which
is a notation reminiscent of the reduced matrix elements of the Wigner-Eckart
theorem. With this definition we finally obtain the expression for
the normalized excited state wave function 
\begin{equation}
|2^{+}M\rangle=\frac{\hat{Z}_{2M}}{\langle||\hat{Z}_{2}^{+}\hat{Z}_{2}||\rangle^{1/2}}|\phi(0)\rangle+O(q_{2M})\label{eq:2m}
\end{equation}
 The wave function $|2^{+}M\rangle$ is well defined in the $q_{2\mu}\rightarrow0$
limit and is a linear combination of 2-quasiparticle excitations
of the spherical state. The expressions in Eqs (\ref{eq:2m})
and (\ref{eq:GS0}) can be now used in the defining formula for
the $B(E2)$ transition strength 
\begin{equation}
B(E2,0^{+}\rightarrow2^{+})=\frac{5}{4\pi}\sum_{M\mu}|\langle2^{+}M|\hat{Q}_{2\mu}^e|0^{+}\rangle|^{2}\label{eq:BE2}
\end{equation}
 where $\hat{Q}_{\lambda\mu}^e$
is the standard electric multipole operator of rank $\lambda$. 
Taking the expressions for the wave functions in the small deformation
limit,  the matrix element becomes
$$
\langle\phi(0)|(\hat{Z}_{2M})^{+}\hat{Q}_{2\mu}^e|\phi(0)\rangle=\delta_{\mu M}\langle||\hat{Z}_{2}^{+}\hat{Q}_{2}^e||\rangle
$$
The final expression for the $B(E2)$ is 
\begin{equation}
B(E2,0^{+}\rightarrow2^{+})_{|\textrm{Sph}}=5\frac{5}{4\pi}\frac{|\langle||\hat{Z}_{2}^{+}\hat{Q}_{2}^e||\rangle|^{2}}{\langle||\hat{Z}_{2}^{+}\hat{Z}_{2}||\rangle}\label{eq:BE2sph}
\end{equation}
The generalization to arbitrary multipolarity $\lambda$
is
\begin{equation}
B(E\lambda,0^{+}\rightarrow\lambda^{\pi_{\lambda}})_{|\textrm{Sph}}
=(2\lambda+1)\frac{2\lambda+1}{4\pi}\frac{| \langle || \hat{Z}_\lambda^+ \hat{Q}_\lambda^e || \rangle |^2}
{ \langle || \hat{Z}_\lambda^+ \hat{Z}_\lambda || \rangle}.
\label{eq:BELsph}
\end{equation}
In contrast to the rotational formula, 
Eq (\ref{eq:BE2sph}) is nonzero in the spherical limit.
This is a clear indication of the inadequacy of the
rotational formula for the evaluation of transition strengths
near sphericity. %This is an important result as the rotational formula
%is routinely used to establish the link between intrinsic deformation
%parameters and the observed transition probabilities. 

The quantities entering Eqs (\ref{eq:BE2sph}) and (\ref{eq:BELsph})
can be calculated in linear response theory, but it is rather easy to 
calculate them using the intrinsic states of the HFB theory.
The only additional computational capability needed is the evaluation
of matrix elements between different intrinsic states.
In particular, we make use of the matrix element of quadrupole operator between 
deformed and spherical states given by 
\begin{equation}
\langle\phi(q_{2\mu})|\hat{Q}_{2\nu}^e|\phi(0)\rangle=
-iq_{2\nu}\langle||\hat{Z}_{2}^{+}\hat{Q}_{2}^e||\rangle+O(q_{2\nu}^{2})
\label{eq:ZQover}.
\end{equation}
To get the normalization in Eq. (7), we make use of the derivatives of 
the overlap function.  The second derivative of the overlap between two
intrinsic wave functions satisfies
\begin{equation}
\gamma=\frac{\partial^{2}}{\partial q_{2\nu}\partial q'_{2\nu'}}
\langle\phi(q_{2\nu})|\phi(q'_{2\nu'})\rangle_{|q_{2}q'_{2}\rightarrow0}=\langle||\hat{Z}_{2}^{+}\hat{Z}_{2}||\rangle\delta_{\nu\nu'}\label{eq:gamma}.
\end{equation}
The second derivative can be approximated
by a finite difference formula in the limit $q_{2\nu}\rightarrow0$
 
\begin{equation}
\gamma=\lim_{q_{2\nu}\rightarrow0}
\frac{\left(\langle\phi(q_{2\nu})|-\langle\phi(-q_{2\nu})|\right)\left(|\phi(q{}_{2\nu})\rangle-|\phi(-q{}_{2\nu})\rangle\right)}
{4q_{2\nu}^{2}}
\label{eq:GammaFD}.
\end{equation}
 Using this result and Eq (\ref{eq:ZQover}) we obtain the following result
for the $B(E2)$
in the spherical limit, 
\begin{equation}
B(E2,0^{+}\rightarrow2^{+})_{|\textrm{Sph}}=5\frac{5}{4\pi}\lim_{q_{2\nu}\rightarrow0}
\frac{|\langle\phi(q_{2\mu})|\hat{Q}_{2\nu}^e|\phi(0)\rangle|^{2}}
{\frac{1}{4}\left(2-\langle\phi(q_{2\nu})|\phi(-q_{2\nu})\rangle-\langle\phi(-q_{2\nu})|\phi(q{}_{2\nu})\rangle\right)}.
\label{eq:BE2SphAO}
\end{equation}
 It is worth remarking that this derivation is valid for any value
of $\nu$ and therefore the axial case corresponding to $\nu=0$ can
be used as well. This formula could be easily implemented in Wood-Saxon
codes to obtain a quick estimate of the spherical transition strength.

If this reasoning is applied to the octupole case, the $|3^{-}M\rangle$
wave function is given by the expression 
\begin{equation}
|3^{-}M\rangle=\frac{\hat{Z}_{3M}}{\langle||\hat{Z}_{3}^{+}\hat{Z}_{3}||\rangle^{1/2}}|\phi(0)\rangle+O(q_{3M})\label{eq:3m}.
\end{equation}
This coincides with the negative parity projected wave function $|\Psi_{-}(q_{3\mu})\rangle=\mathcal{N}_{-}(1-\hat{\Pi})|\phi(q_{3\mu})\rangle$
up to order $q_{3\mu}$. On the other hand, the $|0^{+}\rangle$ wave
function is given by the positive parity projected wave function $|\Psi_{+}(q_{3\mu})\rangle=\mathcal{N}_{+}(1+\hat{\Pi})|\phi(q_{3\mu})\rangle=|\phi(0)\rangle+O(q_{3\mu}^{2})$.
Taking into account these quantities in the general definition of
Eq (\ref{eq:BELsph}) we arrive at the formula
\begin{equation}
\label{eq:BE3Sph}
B(E3,0^{+}\rightarrow3^{-})_{|\textrm{Sph}}
\approx 7\frac{7}{4\pi}|\langle\Psi_{-}(q_{30})|\hat{Q}_{30}^e|\Psi_{+}(0)\rangle|^{2}
\end{equation}
Use of this formula of course requires that the $q_{30}$ in the 
negative parity wave function is small enough so that
this wave function is well approximated by Eq (\ref{eq:3m}).
We used this formula recently in a global study of octupole 
correlations \cite{rob11} to understand some discrepancies 
observed in the comparison with experimental data.

We finish this section by mentioning that the previous methodology can
also be used with scalar operators like the Hamiltonian. It is possible
to obtain in this way formulas for the energies of $J \ne 0$ states in the 
spherical limit. This is briefly discussed in the appendix.

\section{Comparison with exact projected transition strengths}
\subsection{Validity of rotational formula}

\begin{figure}
\includegraphics[width=0.95\columnwidth]{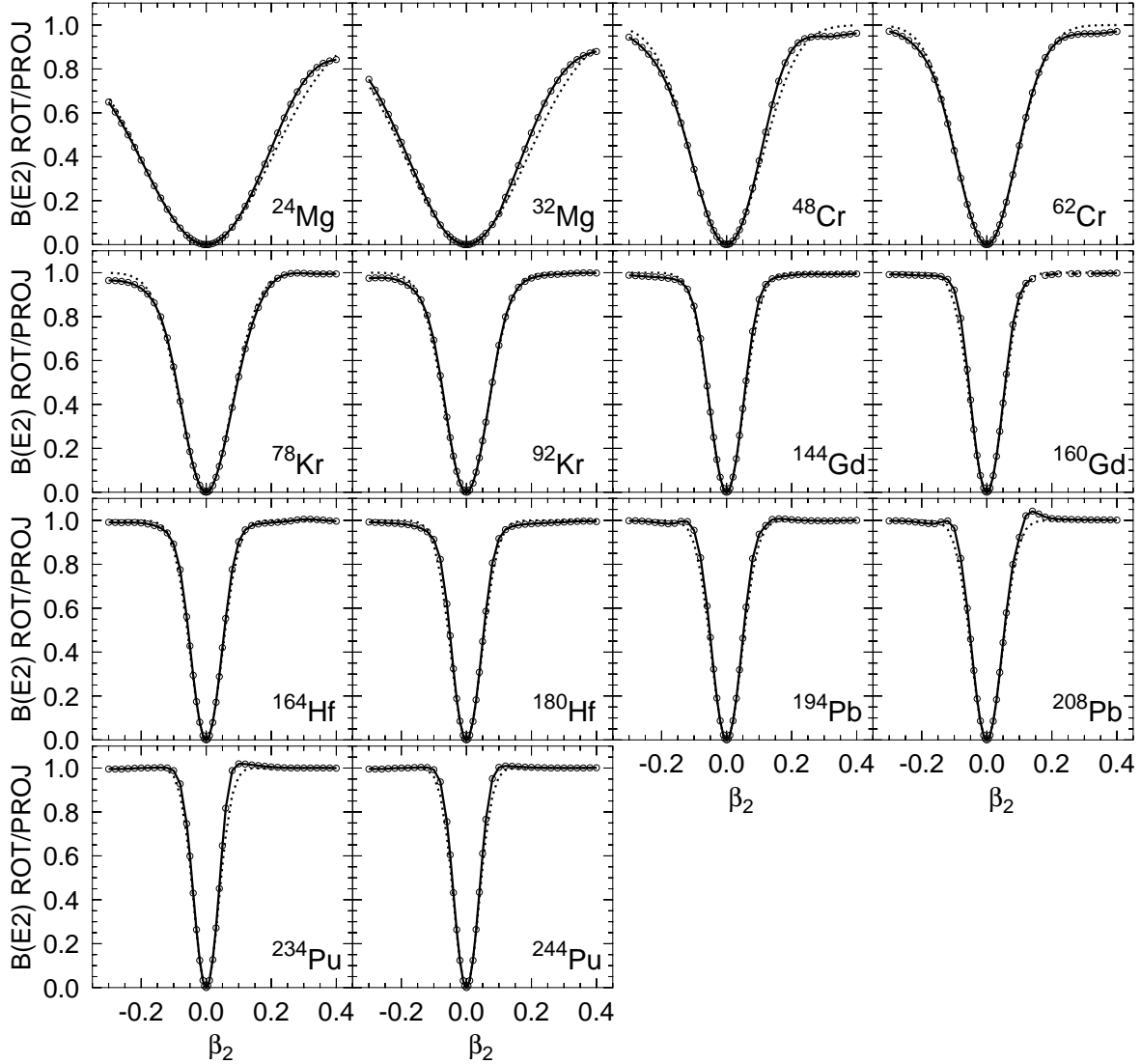}
\caption{The ratio $B(E2)_{\textrm{ROT}}/B(E2)_{\textrm{PROJ}}$ is plotted
as a function of the deformation parameter $\beta_{2}$ for a range of
nuclei.  The solid line connects calculated values. The dashed line 
is calculated from the interpolating formula, Eq. (\ref{eq:BE2ProjAPP}) and 
(\ref{beta0}).
}
\label{fig:BE2Rat_BETA2}
\end{figure}

\begin{figure}
\includegraphics{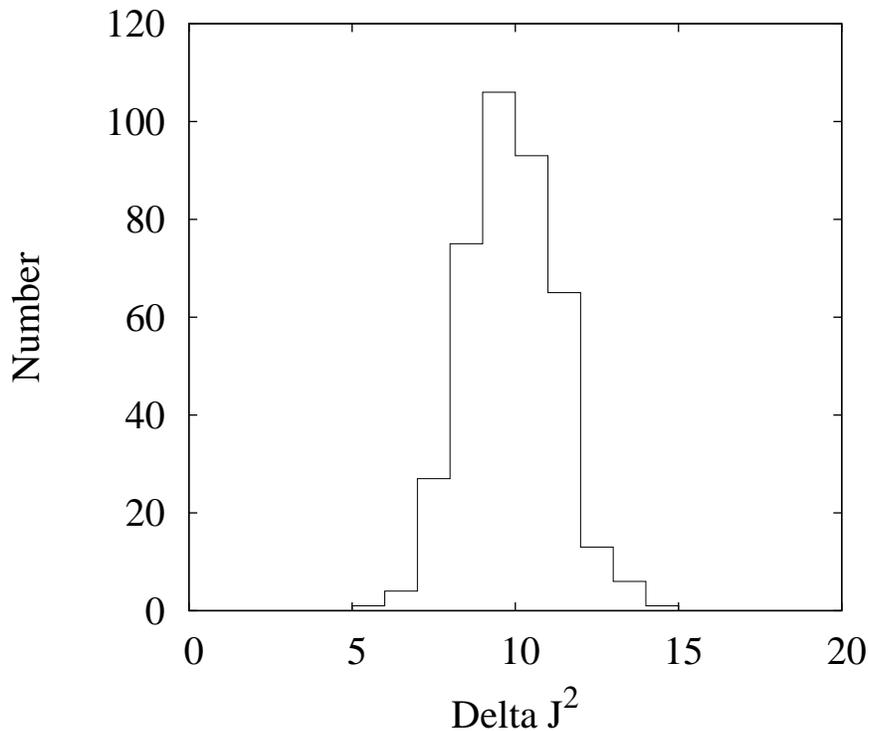}
\caption{Lowest $\langle \Delta \vec{J}^2\rangle$ values of intrinsic wave functions
that meet our criterion
for using the rotational formula (see text). 
\label{fig:djx}}
\end{figure}

In this section we compare the transition strengths computed with
exact angular momentum projection with the rotational formula and 
our spherical limit. The mean-field wave functions were calculated
in the Hartree-Fock-Bogoliubov approximation assuming axial symmetry
and obtaining a range of deformations by including an external 
quadrupole field in the Hamiltonian.  The range of deformations
$\beta_2$ spans the interval $-0.3$ to 0.4 in steps of 0.02 and a finer mesh with
a step size of 0.01 is used in the -0.1 to 0.1 interval\cite{b2}. 
For those intrinsic wave functions the $B(E2,2^{+}\rightarrow0)$ transition
strength has been computed with the rotational formula and exact
angular momentum projection with $|\phi(\beta_{2})\rangle$ as the
intrinsic states (see \cite{ray02} for the relevant formulas). In
Fig. \ref{fig:BE2Rat_BETA2} the ratio $B(E2)_{\textrm{ROT}}/B(E2)_{\textrm{PROJ}}$
is plotted as a function of $\beta_{2}$ for a sample of nuclei
spanning a wide range of masses.  As expected, the ratio increases
toward one as $\beta_{2}$ becomes large.  However, the limit is only 
reached in medium and heavy nuclei within our range of $\beta_2$ values.
For small values of $\beta_{2}$ the ratio is smaller than one and 
approaches zero as $\beta_{2}\rightarrow0$.  One can see that the
$\beta_2$ value by itself does not provide a good indicator of the
region of validity of the rotational formula.  To get a more robust
criterion, we go back to the basic assumption in deriving the
rotational formula, that the intrinsic states have vanishing overlaps
under finite rotations of the orientations. This requires a large
angular momentum content of the intrinsic states.  The mean square
angular momentum of the intrinsic state $\langle \Delta \vec{J}^2\rangle$
can be easily computed from the HFB wave function, so we may consider
that quantity as a practical indicator.  
We note that overlap between rotated wave functions approaches a
Gaussian of width $1/\langle\Delta\vec{J}^{2}\rangle$ \cite{RS}. This result
suggests that the validity of the rotational formula could be linked
to specific values of $\langle\Delta\vec{J}^{2}\rangle$. To explore
this possibility we have determined the value of $\langle\Delta\vec{J}^{2}\rangle$
for the intrinsic configuration that satisfies $B(E2)_{\textrm{PROJ}}/B(E2)_{\textrm{ROT}}\approx3/4$
(a value we have chosen to establish the limits of validity of the
rotational formula) in each of the nuclei of our calculation. The values
are shown as a histogram in Fig. \ref{fig:djx}.  We see that the values are 
strongly peaked around $\langle\Delta\vec{J}^{2}\rangle\approx10\hbar^{2}$.
This remarkable fact gives us an easily  computed estimator of the
validity of the rotational formula for the $B(E2)$ transition strength
for any nucleus in the Chart of Nuclides. %The value is independent of mass
%number which is consistent with the fact that for heavy nuclei a given
%value of $\langle\Delta\vec{J}^{2}\rangle$ is attained at a much
%lower quadrupole deformation than in light nuclei.

\begin{figure}
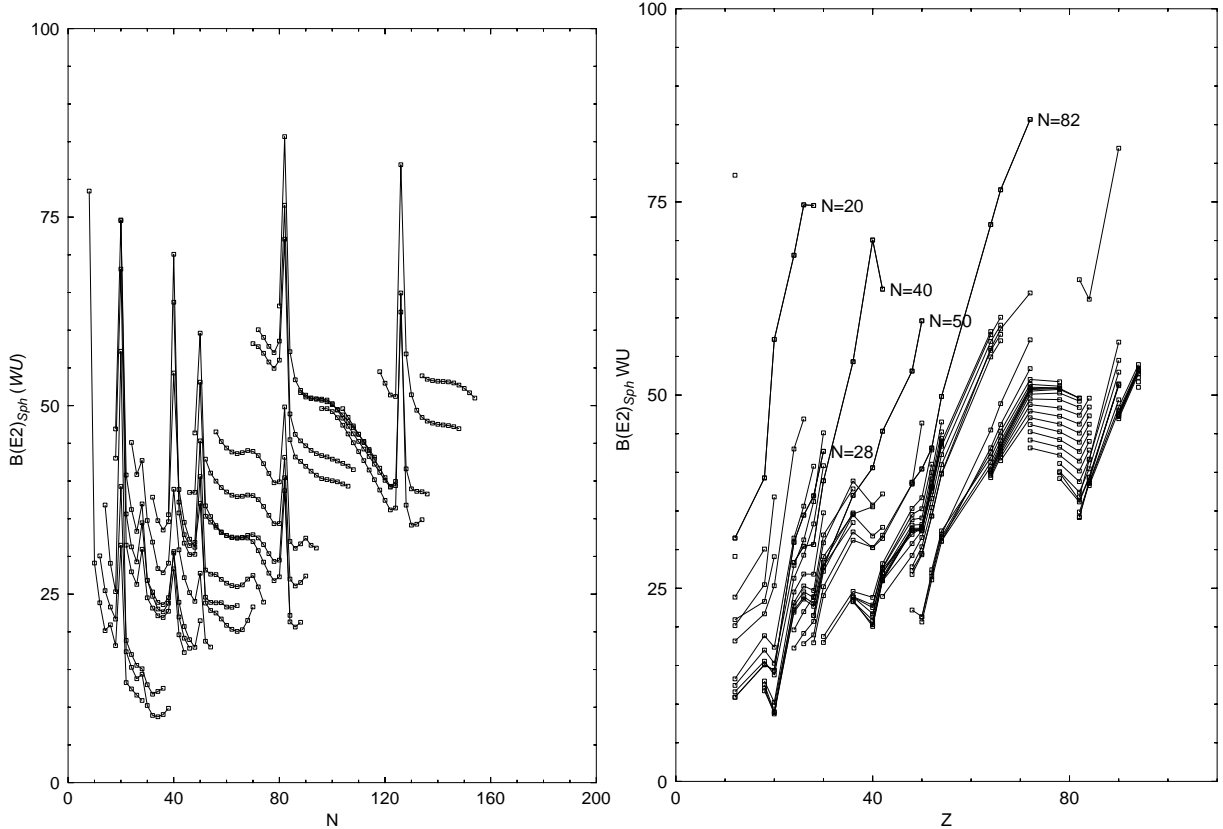

\includegraphics[width=0.49\columnwidth]{BE2Sph}%
\includegraphics[width=0.49\columnwidth]{BE2SphZ}
\caption{The spherical-limit transition strengths of Eq (\ref{eq:BE2sph}) are displayed for several
isotopic chains as a function of neutron number N on the left panel
and as a function of Z in the right panel. The isotopic chains correspond
to Z values between 12 and 94 in steps of 6 units.\label{fig:BE2_ZN} 
Strengths are given in Weisskopf units, 1 W.u. =
$5.94\times10^{-6}A^{4/3}\,\,\,e^2\textrm{b}^{2}$.
}
\end{figure}

\subsection{Selected isotope and isotone chains}

The behavior of the spherical $B(E2)$ transition strengths as
a function of proton and neutron numbers is analyzed next. In Fig
\ref{fig:BE2_ZN} the spherical transition strengths of Eq (\ref{eq:BE2sph})
are plotted as a function of neutron number for several isotopic chains.
They have been computed using the exact angular momentum projected
transition strengths for a deformation of the intrinsic state
of $\beta_{2}=0.005$. The values of those spherical transition strengths
are smaller than the typical values of well deformed nuclei that can
reach a few hundreds of W.u. for heavy nuclei. The decrease with
neutron number is rather weak except around magic neutron numbers
where a marked peak is observed. This is probably a consequence of
the lowering of the level density near magic numbers. Surprisingly,
a peak at the non-magic number $N=40$ is also seen. This
behavior is not observed when the quantity is plotted as a function
of proton number (see right panel). First the spherical transition
strength increases with increasing Z values and a reduction at
those values of Z corresponding to magic numbers is observed, specially
at $Z$=82. The values of $B(E2)_{\textrm{Sph}}$ expressed in W.u.
follow a trend with $Z$ that is consistent with the expected linear
behavior in $Z$ based on the scaling of the mean value of proton's
quadrupole moment (remember that W.u. scale like nuclear radius squared).
A least square fit to the computed values for over two hundred nuclei
yields the rule $B(E2,2\rightarrow0)_{\textrm{Sph}}=0.85\, Z$ (W.u.).

\subsection{An interpolating formula} 

Even better than a criterion for the
validity of Eq. (1) would be an interpolating formula that would also
capture the transition region between spherical and strongly deformed
nuclei.  To this end we consider parameterizing the $B(E2)$ by the function
\begin{equation} 
B(E2,2^{+}\rightarrow0)_{\textrm{Int}}
=\frac{C_{0}}{1-\exp[-(\beta_{2}/\beta_{2}^{(0)})^{2}]}\beta_{2}^{2}
\label{eq:BE2ProjAPP} 
\end{equation} 
The parameter $C_0$ is set to
$C_{0}=(9e^{2})/(80\pi^{2})Z^{2}R_{0}^{4}$ to recover the rotational formula
at large deformation.  The parameter $\beta^{(0)}_2$ is set to a value that
reproduces the spherical limit, 
\begin{equation}
\label{beta0}
\beta_{2}^{(0)\,2}={1\over C_{0}}
B(E2,2^{+}\rightarrow0^{+})_{|\textrm{Sph}}. 
\end{equation}
The results obtained with
Eq (\ref{eq:BE2ProjAPP}) are plotted as dashed lines in Fig
\ref{fig:BE2Rat_BETA2}. Remarkably, for most of the cases and for almost the
whole range of $\beta_{2}$ values both the exact and the approximate results
are indistinguishable. It seems that our model can be used with confidence to
compute $B(E2)$ values provided that the parameter $\beta_{2}^{(0)}$ can be
obtained.  

\subsection{Computing the spherical limit}
\begin{figure}
\includegraphics[width=0.5\columnwidth]{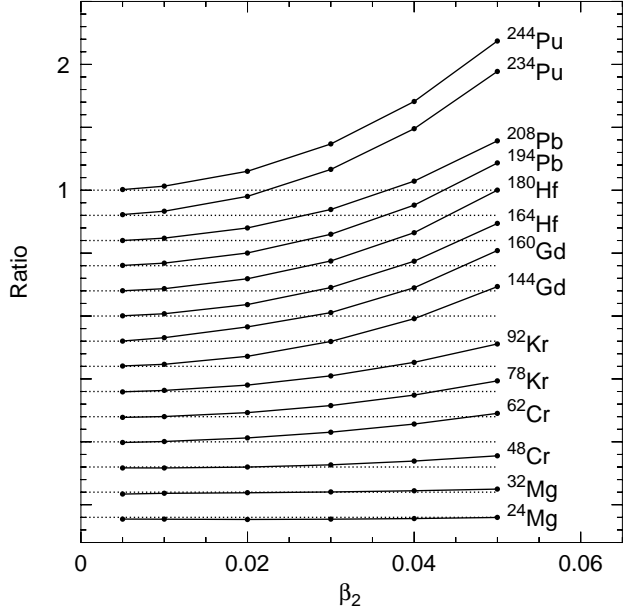} 
\caption{Accuracy of Eq. (\ref{eq:BE2SphAO}) to calculate $B(E2)$ values in the spherical
limit.  Plotted is the ratio of the $B(E2)$ from Eq. (\ref{eq:BE2SphAO}) to the value
obtained by a full projection of the wave functions at $\beta_2=0.005$.
The horizontal scale gives the $\beta_2$ values used in Eq. (\ref{eq:BE2SphAO}).  The
ratios are offset for clarity, with the dotted lines indicating equal
values.\label{fig:Approx}}
\end{figure}

An alternative formula for the evaluation of $B(E2)_{\textrm{Sph}}$ was
obtained in Eq (\ref{eq:BE2SphAO}) in terms of simple overlaps with
the wave functions $|\phi(q{}_{2\nu})\rangle$. To test its applicability
we have performed calculations with our axially symmetric wave functions
as a function of $\beta_{2}$ and some representative results are
given and compared to the exact results in Figure \ref{fig:Approx}.
From the comparison we conclude that the formula is accurate enough
for $\beta_{2}$ values up to 0.05 for light nuclei and up to 0.01
for heavy ones and therefore can be used for a computationally inexpensive
estimation of $B(E2)_{\textrm{Sph}}$ to be used in the model of Eq
(\ref{eq:BE2ProjAPP}) to compute the $\beta_{2}^{(0)}$ parameter
as $\beta_{2}^{(0)}=\left(B(E2)_{\textrm{Sph}}/C_{0}\right)^{1/2}$.

\subsection{Octupole transitions}

\begin{figure}
\includegraphics{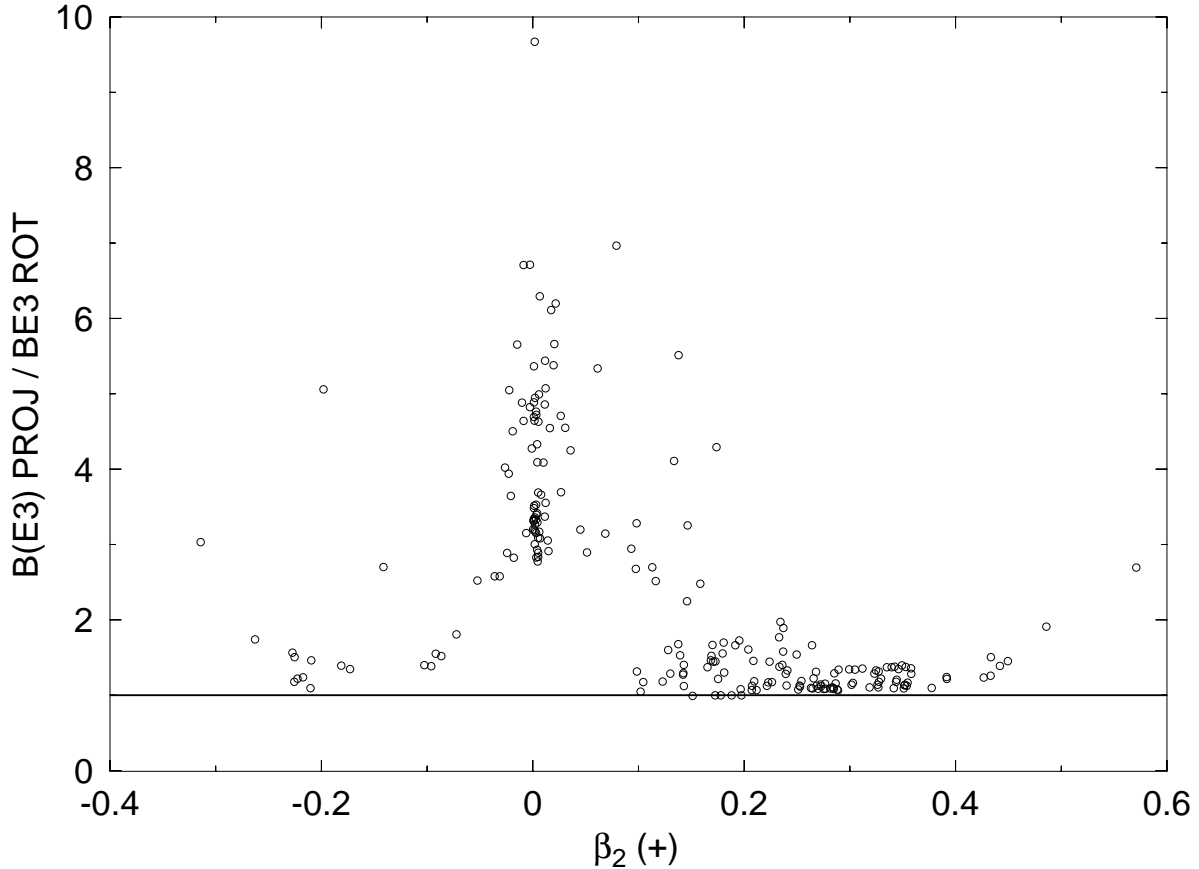}
\caption{The ratio $B(E3,3^{-}\rightarrow0^{+})_{|\textrm{PROJ}}/B(E3,3^{-}\rightarrow0^{+})_{|\textrm{ROT}}$
is plotted as a function of the $\beta_{2}(+)$ quadrupole deformation
parameter of the positive parity intrinsic state.\label{fig:BE3}}
\end{figure}

Another interesting case to study is the one of the $B(E3,3^{-}\rightarrow0)$
transition strengths. They are associated to the octupole degree
of freedom, parameterized in terms of the octupole moments $q_{3\mu}$.
The rotational formula, valid in the strong quadrupole deformation
limit, reads in this case $B(E3,3^{-}\rightarrow0)=\frac{1}{4\pi}|\langle Q_{3}^e\rangle|^{2}$.
Contrary to the quadrupole deformation case, there is no spontaneous
parity symmetry breaking in most of the nuclei of the Nuclide chart
with the exception of a few light Ra and Th isotopes and some rare
earth nuclei like neutron poor Ba isotopes. Therefore the mean value
of the octupole operator in the intrinsic state is zero. As a consequence,
theories dealing with dynamical correlations are required in order
to describe octupole correlations and the associated $B(E3)$. In
those theories the intrinsic octupole deformed state for the $0^{+}$
is different from the one of the $3^{-}$. A typical example is that
of parity projection with restricted variation of the intrinsic state
\cite{rob11}, that assigns the intrinsic states of the $0^{+}$ and
$3^{-}$ states to the ones producing the lowest parity projected
energies $E_{\pm}(q_{3})$ computed for axially symmetric octupole
constrained intrinsic states with octupole deformation $q_{30}$. In
this theory, the rotational formula restricted to axially symmetric
configurations becomes 
$B(E3,3^{-}\rightarrow0^{+})_{|\textrm{ROT}}=
\frac{1}{4\pi}|\langle\Psi_{-}(q_{30}^{(-)})|\hat{Q}_{30}^e|\Psi_{+}(q_{30}^{(+)})\rangle|^{2}$
where now $|\Psi_{\pm}(q_{30})\rangle$ are parity projected wave
functions obtained from an intrinsic state with octupole deformation
$q_{30}$. In order to study the validity of this formula in the spherical
limit, calculations as a function of the quadrupole moment should
be carried out. The difficulty here is that there are two intrinsic
states that potentially have different quadrupole deformations and
therefore a study in terms of four variables (the quadrupole and octupole
moment of the positive and negative parity intrinsic states) should
be carried out for a series of isotopes. Instead of this long calculation
we have just taken the intrinsic states for positive and negative
parity from the results of \cite{rob11} and computed the corresponding
transition strengths with angular momentum projected wave functions.
The ratio $B(E3,3^{-}\rightarrow0^{+})_{|\textrm{PROJ}}/B(E3,3^{-}\rightarrow0^{+})_{|\textrm{ROT}}$
is plotted in Fig \ref{fig:BE3} as a function of the $\beta_{2}(+)$
deformation parameter of the positive parity intrinsic state. Values
corresponding to nuclei where the negative parity quadrupole deformation
parameter $\beta_{2}(-)$ differs significantly from $\beta_{2}(+)$
(by $\pm0.1$) have not been included in the plot. This includes nuclei
with strong shape coexistence where the ground state is, for instance,
prolate and the negative parity state is oblate. As a consequence
of the mismatch in quadrupole deformations the overlap between the
wave functions is very small and the corresponding $B(E3)$ are much
smaller (and therefore more dependent on little details) than for
intrinsic states with similar quadrupole deformation parameters.

The first noteworthy observation is that the transition strengths
computed with the projected angular momentum wave functions are always
greater or equal the values obtained with the rotational formula.
The results show that for $\beta_{2}(+)$ values greater than 0.15
the rotational formula works reasonably well within a factor of 2.
Around $\beta_{2}(+)=0$ the ratio lies in between 3 and 8 in good
agreement with the results of Eq (\ref{eq:BE3Sph}) that predict a
factor 7 difference with the rotational formula in the spherical limit.
The main conclusion is that for quadrupole deformations smaller than
$\beta_{2}\approx0.15$ the rotational formula should not be trusted
and its use avoided in relating transition strengths to intrinsic
octupole deformation parameters. A typical example illustrating the
general trend is that of $^{208}$Pb where the rotational formula
predicts a $B(E3)$ value of 7.1 W.u. whereas the transition strength
with the angular momentum projected wave functions is 23.1 W.u. which
is in much better agreement with the experimental data of 34 W.u..
Such enhancement of the $B(E3)$ transition probabilities for near
spherical configurations as compared to the rotational formula was
already noticed in \cite{egi93,egi96} for some spherical or near
spherical nuclei.

\section{Conclusions}

The validity of the rotational formula for multipole transition strengths
is questioned for near spherical configurations. A general formula
to compute those transitions in terms of intrinsic mean values and/or
overlaps is derived by exploiting the simple structure of angular
momentum projected wave functions in the spherical limit. An enhancement
factor of $2\lambda+1$ for transitions of order $\lambda$ is obtained.
Thorough numerical calculations of $B(E2)$ and $B(E3)$ transition
strengths show the validity of the formulas obtained and establish
criteria of validity for the rotational approximation. For quadrupole
transitions, we proposed a simple model to compute the $B(E2)$
and found that it is quite accurate over the entire range of deformation. 
The model contains two parameters that are fixed from the calculated
transition strengths at the two limits, Eq. (2) and (9).
only one parameter, $\beta_2^{(0)}$ that unfortunately is nucleus dependent. 
%Although
%its dependence with mass and charge numbers is established, it is
%not accurate enough for detailed calculations. In that case, the table
%provided in the accompanying material could be of interest. 
We have
also established a criteria to determine the validity of the rotational
formula that only requires the evaluation of a mean field quantity:
the fluctuation $\langle\Delta\vec{J}^{2}\rangle$ should be larger
than $\sim 10$ for the rotational formula to be useful; it becomes quite
accurate above $\langle\Delta\vec{J}^{2}\rangle > 15$.
For octupole transition strengths $B(E3)$, the quadrupole deformation
parameter $\beta_{2}$ of the ground state has to be larger than $0.15$
for the rotational formula to be valid and precautions are in order
for those cases of shape coexistence where the quadrupole deformation
parameters of positive an negative parity states differ considerably.
For spherical configurations the $B(E3)$ can be up to a factor of
8 larger than the values provided by the rotational formula. 

A table is provided, as supplementary material, with the spherical 
$B(E2)$ strengths and the $\beta_2^{(0)}$ parameters for 818 even-even 
nuclei computed with the Gogny D1S interaction.

\begin{acknowledgments}
This work was supported in part by the U.S. Department of Energy under
Grant DE-FG02-00ER41132, and by the National Science Foundation under
Grant PHY-0835543. The work of LMR was supported by MINECO (Spain)
under grants Nos. FPA2009-08958, and FIS2009-07277, as well as by
Consolider-Ingenio 2010 Program MULTIDARK CSD2009-00064. 
\end{acknowledgments}

\appendix

\section{Projected energies in the spherical limit \label{ENERSph}}

The same arguments used in the previous section can be used to compute
the energy of the $|2^{+}M\rangle$ as given by Eq (\ref{eq:2m})
in the spherical limit 
$$
E(2^{+})_{|\textrm{Sph}}=\frac{ \langle || \hat{Z}_2^+ \hat{H} \hat{Z}_2 || \rangle }{ \langle || \hat{Z}_2^+ \hat{Z}_2 || \rangle } + O (q_{2\mu}^2).
$$
Defining 
$$
h_{qq'}=\frac{\partial^{2}}{\partial q_{2\nu}\partial q'_{2\nu'}}\langle\phi(q_{2\mu})|\hat{H}|\phi(q'_{2\mu'})\rangle_{|q_{2}q'_{2}\rightarrow0}=\langle||\hat{Z}_2^+\hat{H}\hat{Z}_2||\rangle\delta_{\nu\nu'}
$$ 
as using Eq (\ref{eq:gamma}) the excitation energy can be written as 
$$
E(2^{+})_{|\textrm{Sph}}=\frac{h_{qq'}}{\gamma}
$$
an expression that coincides with twice the zero point energy correction
obtained in the Generator Coordinate Method (GCM) for the quadrupole
coordinate in the harmonic limit of the Gaussian Overlap Approximation
(GOA) (See Eq (10.136) of \cite{RS}). The energy of the $2^{+}$
state in the spherical limit is not given in calculations with
angular momentum projection as its evaluation involves the ratio of
two very small quantities which are difficult to compute with the
required accuracy \cite{ray02}.

% ------------------------------------------------------------------------
%                       B I B L I O G R A P H Y 
% ------------------------------------------------------------------------

%

\begin{thebibliography}{10}
\bibitem{Qdef} Note that this differs by a factor of $\sqrt{2 \lambda + 1 \over 4
\pi}$ from the definition in \cite{BM}.
\bibitem{b2}  We use the standard practical definition of $\beta_2$,
related to the mass quadrupole moment $Q_2$ by 
$Q_2=\sqrt{\frac{4\pi}{5}}\frac{3}{4\pi}AR_{0}^{2}\beta_{2}$
with $R_{0}=1.2A^{1/3}$ fm.
\end{thebibliography}
\end{document}